\documentclass {aastex63}
\usepackage{comment, subfigure}
\usepackage{mathrsfs}
 \turnoffedit
\usepackage{amsfonts}
\usepackage{mathtools}
\usepackage{booktabs}
\usepackage[section]{placeins}

\begin{document}

\shorttitle{Double-Degenerate Near-$M_{\rm Ch}$ SNe Ia}
\shortauthors{Neopane and Bhargava et al.}

\title{Near-Chandrasekhar-Mass Type Ia Supernovae from the Double-Degenerate Channel}

\author[0000-0003-3944-3645]{Sudarshan Neopane}
\altaffiliation{These authors contributed equally to this work.}
\affiliation{Department of Physics, University of Massachusetts Dartmouth, 285 Old Westport Road, North Dartmouth, MA 02747, USA}

\author[0000-0003-0385-7918]{Khanak Bhargava}
\altaffiliation{These authors contributed equally to this work.}
\affiliation{Department of Physics, University of Massachusetts Dartmouth, 285 Old Westport Road, North Dartmouth, MA 02747, USA}

\author[0000-0001-8077-7255]{Robert Fisher}
\email{rfisher1@umassd.edu}
\affiliation{Department of Physics, University of Massachusetts Dartmouth, 285 Old Westport Road, North Dartmouth, MA 02747, USA}

\author[0000-0003-4203-4973]{Mckenzie Ferrari}
\affiliation{Department of Physics, University of Massachusetts Dartmouth, 285 Old Westport Road, North Dartmouth, MA 02747, USA}

\author[0000-0002-4641-2751]{Shin’ichirou Yoshida}
\affiliation{Department of Earth Science and Astronomy, Graduate School of Arts and Sciences, The University of Tokyo,Komaba 3-8-1, Meguro-ku, Tokyo 153-8902, Japan}

\author[0000-0002-2998-7940]{Silvia Toonen}
\affiliation{Anton Pannekoek Institute, University of Amsterdam, Science Park 904, 1098 XH Amsterdam, Netherlands}

\author[0000-0003-0894-6450]{Eduardo Bravo}
\affiliation{E.T.S. Arquitectura del Valle$\grave{s}$, Universitat Polite$\grave{c}$nica de Catalunya, Carrer Pere Serra 1-15, 08173 Sant Cugat del Valle$\grave{s}$, Spain}

\correspondingauthor{Robert Fisher}




\begin{abstract}

Recent observational evidence has demonstrated that white dwarf (WD) mergers are a highly efficient mechanism for mass accretion onto WDs in the galaxy. In this paper, we show that WD mergers naturally produce highly-magnetized, uniformly-rotating WDs, including a substantial population within a narrow mass range close to the Chandrasekhar mass ($M_{\rm Ch}$). These near-$M_{\rm Ch}$ WD mergers subsequently undergo rapid spin up  and compression on a $\sim 10^2$ yr timescale, either leading to central ignition and a normal SN Ia via the DDT mechanism, or alternatively to a failed detonation and SN Iax through pure deflagration. The resulting SNe Ia and SNe Iax will have spectra, light curves, polarimetry, and nucleosynthetic yields similar to those predicted to arise through the canonical near-$M_{\rm Ch}$ single degenerate (SD) channel, but with a $t^{-1}$ delay time distribution characteristic of the double-degenerate (DD) channel. Furthermore, in contrast to the SD channel,  WD merger  near-$M_{\rm Ch}$ SNe Ia and SNe Iax will not produce observable companion signatures.  We discuss a range of implications of these findings, from SNe Ia explosion mechanisms, to galactic nucleosynthesis of iron peak elements including manganese.

\end{abstract}


\received {October 18, 2021}
\accepted {November 17, 2021}
\submitjournal{The Astrophysical Journal}


\section{Introduction}
\label{sec:intro}

Type Ia supernovae (SNe Ia) are the radioactively-powered thermonuclear explosions of white dwarfs (WDs), and play a crucial role as standardizable candles for cosmology \citep {pankey62, phillips93}. The stellar progenitors of SNe Ia have remained a subject of intense investigation for decades. For many years, the single-degenerate (SD) channel, in which a main sequence or red giant star donates material to a CO white dwarf until it reaches a near-Chandrasekhar mass $M_{\rm Ch}$ and subsequently detonates as a SN Ia \citep {whelaniben73}, was considered canonical. More recent evidence has begun to favor the double-degenerate (DD) SNe Ia channel, resulting from the merger of two white dwarfs \citep {webbink84}. However, numerous questions still surround both channels, leaving the nature of the SNe Ia progenitors a largely unsolved problem \citep {maozetal13}.

A key outstanding puzzle surrounding SNe Ia  stems from the nucleosynthesis of iron group elements (IGEs). The fundamental physics of electron degeneracy requires that the central density of WDs increases with increasing WD mass \citep {chandrasekhar35}. Consequently, near-$M_{\rm Ch}$ WD progenitors undergo more efficient electron capture in their cores than sub-$M_{\rm Ch}$ progenitors. This efficient electron capture in near-$M_{\rm Ch}$ WD progenitors in turn results in enhanced abundances of neutron-rich isotopes, most notably the radiosotope $^{55}$Fe, which decays to the monoisotopic $^{55}$Mn \citep {ropkeetal12}. \citet {seitenzahletal13} compellingly argue that a substantial fraction of near-$M_{\rm Ch}$ WD events, as much as half of the total SNe Ia rate, is required to produce a solar abundance of $^{55}$Mn. Detailed galactic nucleosynthesis models find the fraction of near-$M_{\rm Ch}$ SNe Ia to be as high as 75\% in the solar neighborhood, though lower in dwarf galaxies \citep {kobayashietal19}. 

Another puzzle surrounding near-$M_{\rm Ch}$ events is connected to subluminous SNe Iax. SNe Iax are relatively commonplace transients, accounting for 10- 30\% of the total SNe Ia rate \citep {foleyetal13}. The leading theoretical models for SNe Iax are failed detonations of near-$M_{\rm Ch}$ WDs \citep{jordanetal12, finketal14}. Yet, the rates for the production of near-$M_{\rm Ch}$ WDs predicted from binary population synthesis (BPS) models through the canonical SD channel is typically an order of magnitude less than the total SNe Ia rate, and depending on model assumptions, orders of magnitude less \citep {maozetal14}. Thus there is also a tension between the requisite SNe Iax rate and that predicted from BPS models, if SNe Iax originate as  failed detonations of near-$M_{\rm Ch}$ WDs through the classical SD channel.

Furthermore, numerous observations place tight constraints on the prevalence of near-$M_{\rm Ch}$ SNe Ia events produced through the canonical single-degenerate channel. Observations of the nebular phase spectra of a large sample of 111 SNe Ia including normal, 91T-like, and 91bg-like (but excluding Ia-CSM and Iax) events found no evidence of H${\alpha}$ from the stripped companions expected from the SD channel \citep {tuckeretal19}, contrary to theoretical expectations \citep {botyanszkietal18}. Similarly tight limits on single-degenerate SNe Ia are obtained from companion \citep {lietal11, shappeeetal16b},  and ex-companion \citep {gonzalezhernandez12, schaeferpagnotta12} searches, the delay-time distribution (DTD) \citep {strolgeretal20}, as well as radio \citep {chomiuketal16} and X-ray \citep {marguttietal12, marguttietal14} constraints. 

In this paper, we present a possible solution to the challenges posed by  near-$M_{\rm Ch}$ SNe Ia and SNe Iax. We propose that near-$M_{\rm Ch}$ SNe Ia and SNe Iax are produced primarily not through the SD channel, as is widely believed, but instead through the DD channel. Our line of reasoning builds upon recent observational advances,  showing  that the galactic WD merger rate exceeds the SN Ia rate  \citep {maozetal18, chengetal19}, and that at least  half of the most massive WDs in the solar neighbourhood are merger products  \citep{kilic2021}.

The production of super-$M_{\rm Ch}$ WD mergers through the DD channel exceeds the predicted SD channel rate, provided that only a small fraction of super-$M_{\rm Ch}$ mergers promptly detonate upon merger, as both recent hydrodynamical simulations \citep {raskinetal12, raskinetal14, kashyapetal17} and observations \citep {shenetal18} suggest. Additional theoretical and observational investigations have demonstrated that the product of the merger is a high-field magnetic white dwarf (HFMWD) \edit1{ \citep {jietal13, Gvaramadze_2019, caiazzo2021}}. We will show that a substantial population of mergers result in rigidly-rotating WDs in a  narrow mass range close to $M_{\rm Ch}$ \edit1 {which} lose angular momentum through both a propeller-driven phase as well as magnetic dipole radiation. These mergers have been previously shown to  undergo central ignition \citep {becerraetal18, becerraetal19}. We will  consider both the possibility that these central ignitions lead to SNe Ia as well as failed detonation SNe Iax, and explore a SN Ia model in detail.

The structure of this paper is as follows. In \S 2, we demonstrate how WD mergers are naturally expected to give rise to near-$M_{\rm Ch}$ SNe Ia and SNe Iax progenitors. In \S 3, we describe the methodology used to numerically simulate a near-$M_{\rm Ch}$ DD SNe Ia event, including the calculation of its detailed nucleosynthetic abundances and synthetic spectra. In \S 4, we present the simulation results. In \S 5, we discuss the observational and theoretical implications of  WD merger near-$M_{\rm Ch}$ SNe Ia and SNe Iax. Finally, in \S 6, we conclude.

\section{Near-Chandrasekhar Mass SNe Ia From DD WD Mergers} 
\label{sec:Iadd}

\subsection {Post-Merger HFMWD Production} \label {sec:hfmwd}

Decades of theoretical work have elucidated the physical processes at work during WD mergers. WDs in tight binaries formed through previous common envelope phases of evolution \citep {nelemanstout05, distefano10} are driven together in a final inspiral by gravitational wave radiation. As the secondary WD approaches the tidal radius, it is tidally disrupted, producing a hot, flared accretion disk surrounding the primary WD \citep {benzetal90, lorenaguilaretal09}. The differential rotation within the disk is unstable to the magnetorotational instability \citep {balbushawley91}, so that even a dynamically weak seed field is rapidly amplified over several disk rotational periods \citep {jietal13}. Consequently, magnetohydrodynamics plays a crucial role both in regulating the accretion and outflow rates, and in determining the final thermodynamic structure and spin of the WD merger. 

\begin{figure*}
\gridline{\fig{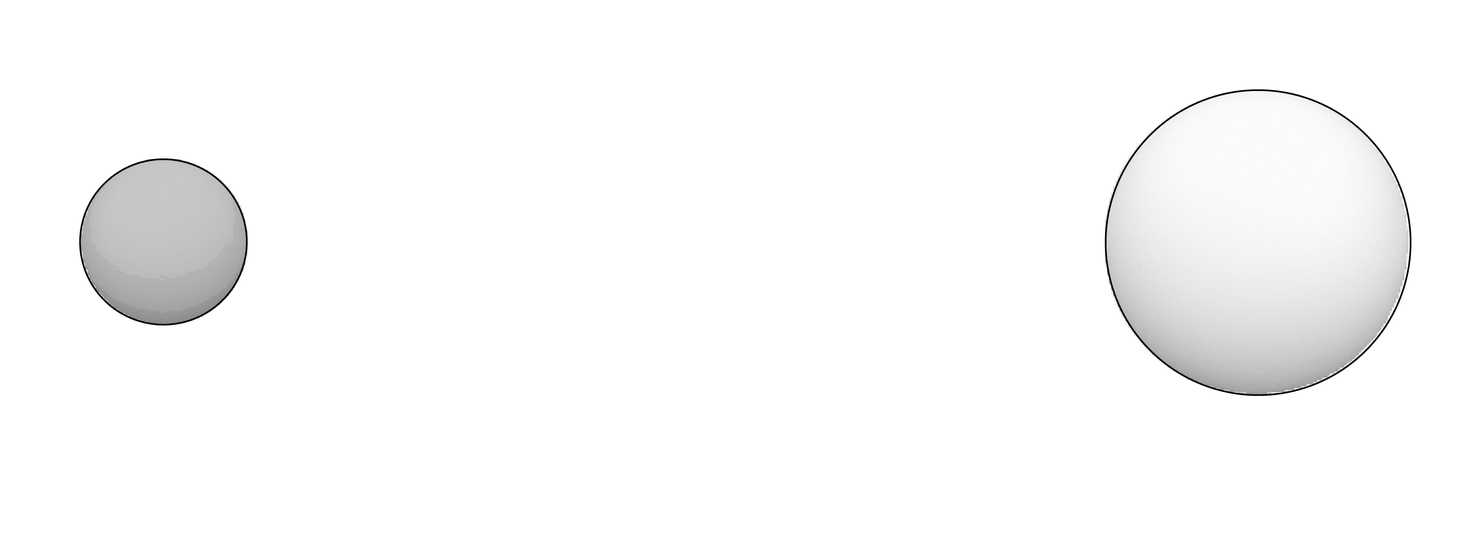}{0.25\textwidth}{(a)}
          }
\gridline{\fig{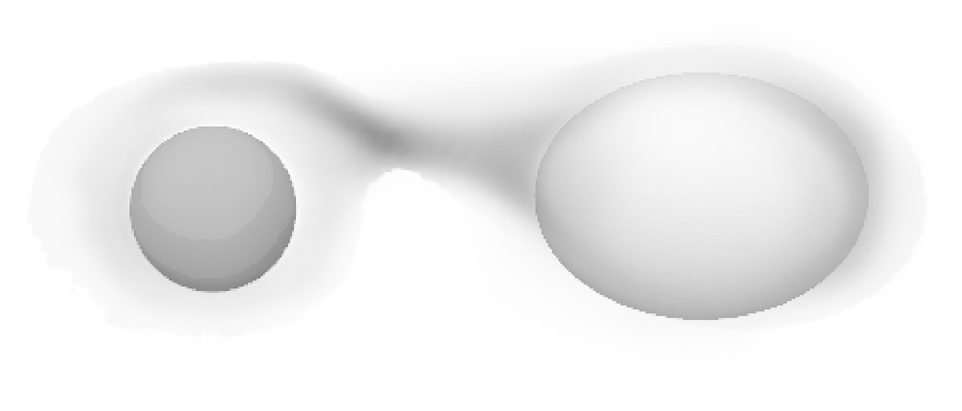}{0.2\textwidth}{(b)}
          }
\gridline{\fig{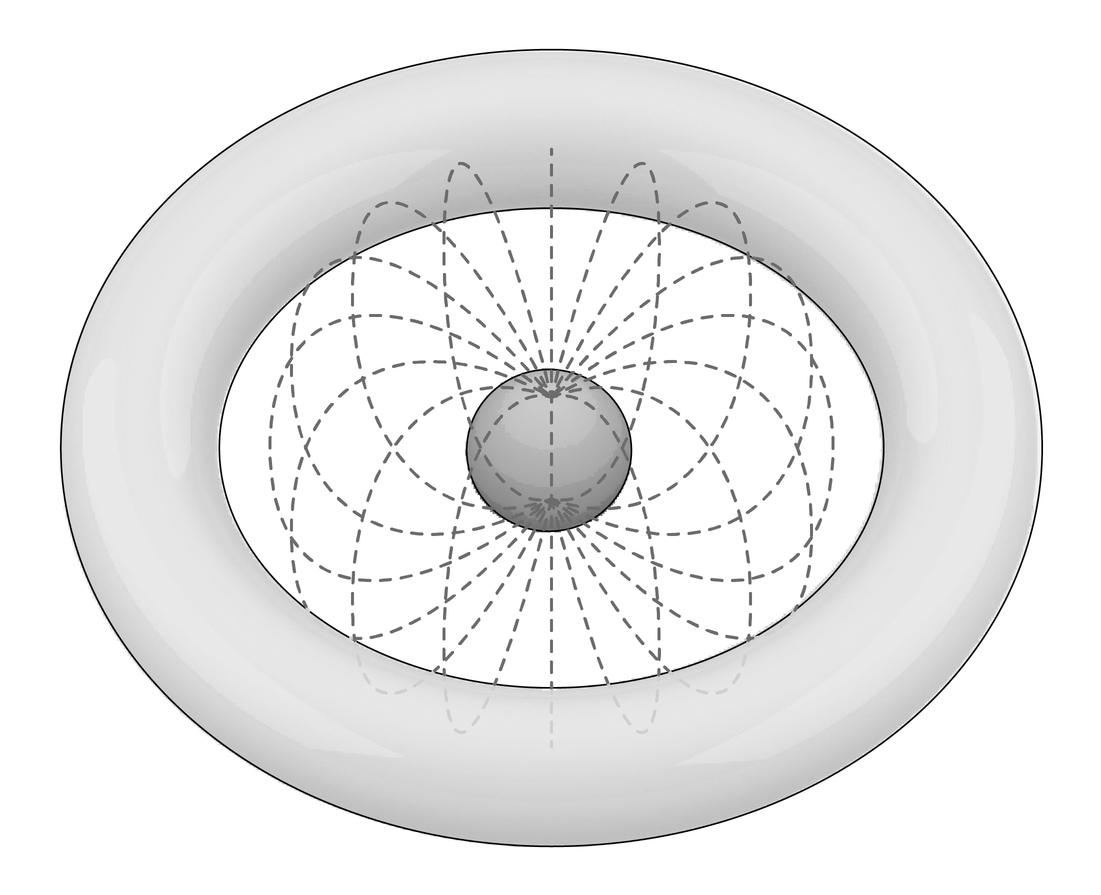}{0.2\textwidth}{(c)}
          }
\gridline{\fig{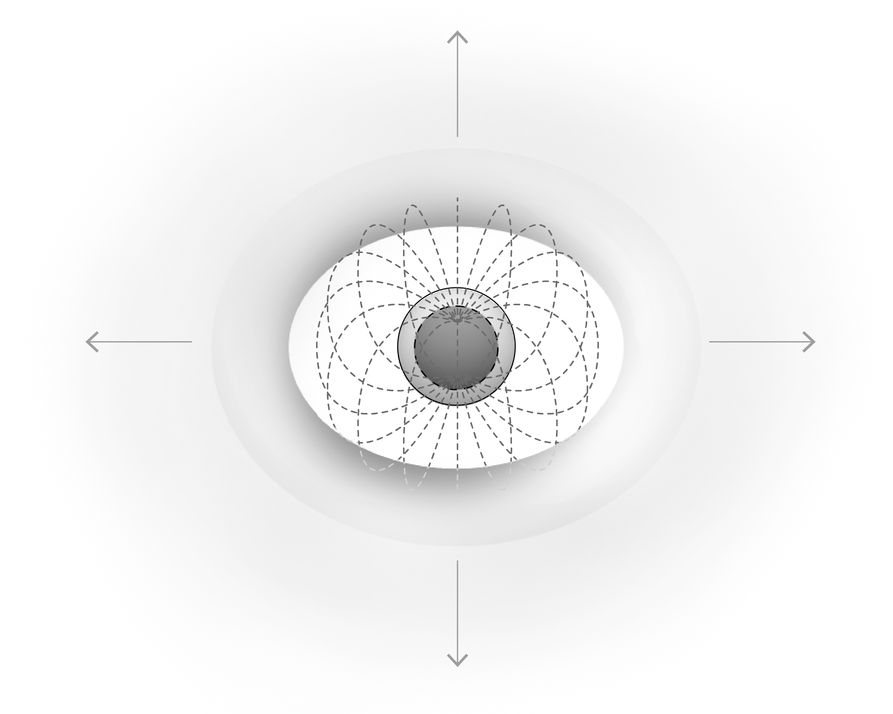}{0.3\textwidth}{(d)}
          }
\gridline{\fig{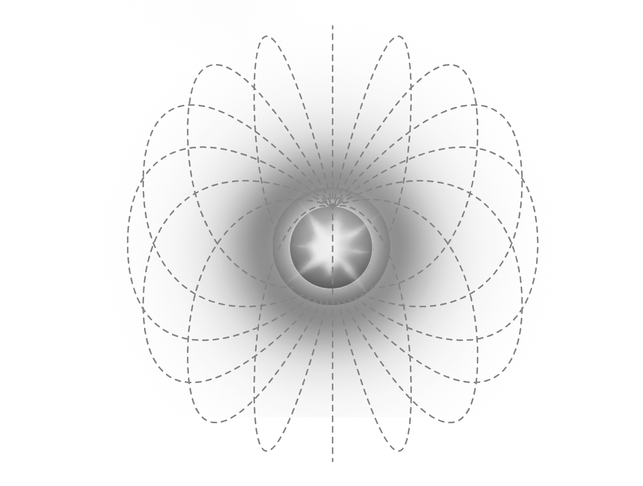}{0.2\textwidth}{(e)}
          }
\caption{Schematic diagram showing the processes involved in the formation of near-$M_{\rm Ch}$ SNe Ia from a DD WD merger. (a) The inspiral phase of the WD merger  lasts $\sim$ 1 - 10\ Gyr. (b) The inspiral phase is followed by the tidal disruption of the secondary on a timescale of $\sim 10^{3}$s. (c) The differentially-rotating disk gives rise to the magnetorotational instability (MRI), the accretion of which onto the primary produces a rigidly-rotating, high field magnetic white dwarf (HFMWD). (d) As the primary accretes from the disk, it gains angular momentum and spins up. On a timescale of $\sim 10^{6}$s, the system enters the propeller phase. During the propeller phase, the primary ejects the remaining disk from the system and spins down. (e) The misaligned magnetic and rotational axes of the the merger cause it to spin up on a timescale of $\sim 10^{2}$yr and compress the core, resulting in central ignition, and either a SN Ia or SN Iax.}
\label{fig:schematic}
\end{figure*}

\subsection {Accretion onto Magnetically-Braked HFMWD} \label {sec:braking}


During the merger itself, material accreted from the tidally-disrupted accretion disk will initially possess a higher specific angular momentum than the primary WD. As a result, the accretion initially generates a differential shear within the WD merger. However, the differential shear within the nascent HFMWD results in the propagation of  torsional  Alfv\'en waves, which magnetically brake the differential rotation of the HFMWD merger rapidly on a timescale of minutes  \citep {shapiro00, piro08},

\begin {equation}
t_{\rm A} = {R \over v_A} \approx 7 \times 10^2 \left(B_0 \over 10^{10}\ {\rm G} \right)^{-1} \left (R_{\rm WD} \over 2 \times 10^3\  {\rm km}  \right)^{-1/2} \left (M \over 1.4\ M_{\odot} \right)^{1/2} {\rm s}
\end {equation}
Here $M$ and $R_{\rm WD}$ are the radius and mass of the HFMWD merger, and $B_0$ is the surface magnetic field strength. $v_A$ is the Alfv\'en velocity $B / \sqrt {4 \pi \rho}$. The Alfv\'enic timescale $t_{\rm A}$ is shorter than the timescale to accrete the tidally-disrupted disk,

\begin {equation}
t_{\rm acc} =  \alpha^{-1} t_{\rm dyn} \approx 3 \times 10^3 \left (0.01 \over \alpha  \right) \left (R_{\rm disk} \over 5 \times 10^4\  {\rm km}  \right)^{3/2}  \left (M \over 1.4\ M_{\odot} \right)^{-1/2} {\rm s}
\end {equation}
Where $t_{\rm dyn}$ is the dynamical timescale corresponding to the outer disk radius $R_{\rm disk}$.  Consequently, the HFMWD is rapidly braked, and is in a state of uniform rotation even as it continues to accrete mass from the disk. This conclusion is confirmed by multidimensional MHD simulations of post-merger accretion \citep {jietal13}, where the resulting HFMWD is found to be uniformly rotating.

Magnetohydroynamics is also responsible for converting differential shear into magnetic energy through the magnetorotational (MRI) instability, and in driving jets and outflows. As such, magnetohydrodynamics plays a leading role in establishing the thermodynamic profile of the disk, and consequently  key observable properties of the merger, including its final radius and surface temperature.  The energy budget of the disk is particularly important because the disk is non-degenerate, and net heating injected into it can result in expanded merger sizes. In particular, if the deposition of differential shear is converted into heat over a viscous timescale through the Shakura-Sunyaev $\alpha$ prescription, the merger expands to $\sim 10^{10} - 10^{11}$ cm \citep {shenetal12, schwabetal12}. In contrast, magnetohydrodynamics converts shear energy into magnetic energy, and propels a significant amount of energy into jets and winds, producing  WD-sized mergers with $\sim  10^{9}$ cm  envelopes \citep {jietal13, becerraetal18}.



\subsection {Propeller Regime} \label {sec:propeller}

If the central WD is rotating sufficiently rapidly, then accretion from the disk onto the WD cannot overcome its centrifugal barrier, and the system enters into the propeller regime.  Theory and simulations demonstrate that accretion efficiency in the propeller regime is greatly reduced. Instead, in the propeller regime, disk material is efficiently ejected into unbound trajectories through magnetically-driven disk winds powered by the rotation of the central star and torquing it down \citep {illarionovsunyaev75, ghoshlamb79, lovelaceetal99, romanovaetal05}. 

We now demonstrate that {\it every double-degenerate white dwarf merger necessarily enters into the propeller regime at some juncture in the accretion process}, provided solely that it does not first detonate promptly as a SN Ia. \citet {ghoshlamb79} defined the fastness parameter $\omega_*$ of the central WD as 
 
 \begin {equation}
\omega_* = {\Omega_{\rm WD} \over \Omega_K (R_{\rm in} ) } = \left (R_{\rm in} \over R_{\rm co} \right)^{3/2}
\end {equation}
Here $R_{\rm co}$ is the corotation radius $R_{\rm CO} = (G M_1 / \omega_{\rm WD}^2)^{1/3}$. When $\omega_* > 1$, the disk is in the propeller regime. One can estimate the inner disk radius $R_{\rm in}$ as a fraction $\xi$ of the Alfv\'en radius, where the inflow velocity of spherical accretion equals the Alfv\'en velocity, such that $R_{\rm in} = \xi R_{\rm A}$,

\begin {equation}
R_{\rm A} = \left ({\mu_*^2 \over \sqrt {2 G M} \dot {M} } \right)^{2/7}
\end {equation}
Here $\mu_*$ is the stellar magnetic dipole moment, and $\dot {M}$ is the mass accretion rate. The propeller regime condition is then simply expressed as $\Omega _{\rm WD} > \Omega_K (R_{\rm in})$, or

\begin {equation}
\Omega_{\rm WD} > \left (G M_{\rm WD} \over \xi^3 R_A^3 \right)^{1/2}
\end {equation}

Solving for $\dot {M}$, the propeller regime condition is

\begin{equation}
\dot{M} < \dot{M}_{\rm crit} = {\xi^{7/2}B_s^2 R_{\rm WD}^6 \over \sqrt {2}\ G^{5/3} M_{\rm WD}^{5/3}} \Omega_{\rm WD}^{7/3}  =  {\xi^{7/2}R_{\rm WD} \over  \sqrt {2}\ \Omega_{\rm max}} \left({\Omega_{\rm WD} \over \Omega_{\rm max} } \right)^{7/3} B_s^2 = {\xi^{7/2} \over 2^{9/2} \pi^2} {\Phi_s^2 \over R_{\rm WD}^2 v_{\rm max}}  \left({ \Omega_{\rm WD} \over \Omega_{\rm max} } \right)^{7/3},
\label{eqn:propeller_regime}
\end{equation}
where $v_{\rm max} = \Omega_{\rm max} R_{\rm WD}$ is the breakup velocity of the WD, $\Omega_{\rm max} = \sqrt {G M_{\rm WD} /R_{\rm WD}^3}$ is the breakup angular velocity of the WD, and $\Phi_s = 4 \pi R_{\rm WD}^2 B_s$ is the mean surface magnetic flux.  

The rotation rate $\Omega_{\rm WD}$ of a rigidly-rotating WD cannot exceed the maximal value, $\Omega_{\rm max} = \sqrt {G M_{\rm WD} /R_{\rm WD}^3}$. Consequently, there is a minimum accretion rate $\dot {M}_{\rm crit}$  crossed when the mass of the accretion disk falls beneath a minimum value, $M_{\rm disk} \sim \dot {M}_{\rm crit} t_{\rm acc}$, below which the system must enter into the propeller regime. The propeller regime has important ramifications for double-degenerate systems. In particular, as a result of the propeller regime, {\it the entire mass of the donor cannot be efficiently accreted onto the WD merger, and there is a minimum amount of mass, of order $\dot {M}_{\rm crit} t_{\rm acc}$, which must be ejected from the system.} We note that the propelled mass is in addition to both the tidal tail \citep {benzetal90, raskinkasen13} and magnetically-driven jets and outflows \citep {jietal13} arising during the earlier phases of the merger; combined this earlier ejected mass amounts to $\sim$ few $\times 10^{-3} M_{\odot}$.

It is important to note that the critical mass accretion rate $\dot {M}_{\rm crit}$ is in turn set by the surface magnetic field of the merger. The propelled mass is clearly a magnetohydrodynamical phenomenon which is absent from a purely hydrodynamical treatment of the merger.


An upper bound estimate of the mass ejected in the propeller phase can be obtained by considering $\dot{M}_{\rm crit}$ and the spin-down timescale, defined as $t_{\rm sd} \approx \Omega_{\rm WD} /|\dot{\Omega}|$.  The spin-down of the WD merger in the propeller phase is due to the magnetic torque acting at the interface of magnetosphere and the inner radius of accretion disk; simulations of the MRI show that the Reynolds stress is negligible in comparison to the Maxwell stress \citep{Davis_2010}. The spin-down of the WD can then be calculated by integrating the magnetic torque over the inner edge of the disk, assuming that the field at the inner edge of the disk is dominated by a dipolar HFMWD field threading the inner disk during the propeller phase, and the approximate equality of the radial and toroidal components of the magnetic field, $B_{r} \approx B_{\phi}$ \citep{Romanova_2003}, 

\begin{equation}
    I \frac{d \Omega}{dt} = -4\pi R_{\rm in}^2 f \left(R_{\rm in} \frac{B_{R_{\rm in}}^2}{8\pi}\right)
\end{equation}

Then, the spin-down timescale is given by
\begin{equation}
     t_{\rm sd} \approx \frac{\Omega_{\rm WD}}{|\dot{\Omega}|} = 2^{4/7} \beta \xi^{3} f^{-1} G^{1/14}    B_{\rm s}^{-2/7} \dot{M}^{-6/7} M_{\rm WD}^{15/14}  R_{\rm WD}^{-5/14} \left( \Omega_{\rm WD} \over \Omega_{\rm max} \right)
\end{equation}
Here $\beta = I/(MR^2)$, where $I$ is the moment of inertia of the WD and $f$ is the fraction of the full spherical solid angle over which matter is being propelled.

An upper bound estimate for the mass ejected from the system $M_{\rm ej}$ can then be obtained by the product of the critical mass accretion rate $\dot {M}_{\rm crit}$ and spin-down timescale $t_{\rm sd}$,

\begin{equation}
   M_{\rm ej} < \dot{M}_{\rm crit}t_{\rm sd} = 2^{1/2} \beta \xi^{7/2} f^{-1}
M_{\rm WD}  \left(\frac{\Omega_{\rm WD}}{\Omega_{\rm max}}\right)^{4/3} \approx 0.2 M_{\odot} \ \left(\frac{\beta}{0.1}\right) \ \left(\frac{\xi}{1.0}\right)^{7/2} \ \left(\frac{f}{1.0}\right)^{-1}   \left(\frac{M_{\rm WD}}{1.4 \ M_{\odot}}\right) \  \ \left(\frac{\Omega_{\rm WD}}{\Omega_{\rm max}}\right)^{4/3} 
\label {eqn:mass_ej}
\end{equation}

Physically, this upper-bound estimate on the propelled mass $M_{\rm ej}$ also follows immediately from the conversion of rotational energy of the WD into propelled mass being ejected at the escape speed at the Alfv\'en radius. Consequently, because the rotational energy of the WD is the ultimate source of energy of the propeller, this upper bound is independent of the magnetic field strength. It is also important to note that the upper bound to the ejected mass $M_{\rm ej}$  is also proportional to the rotational rate of the WD to the $4/3$ power; mergers rotating more slowly than breakup will propel less mass.

\subsection {Central Ignition} \label {sec:ignition}

An off-centered ignition burns through the WD envelope to its center on a timescale of order $t_{\rm burn} \sim 10^4$~yr \citep {nomotoiben85, saionomoto85, schwabetal16}. The inwardly-propagating carbon burning flame initiates a convective zone \citep{schwabetal16}, but continues to stably progress inwards;  \citet{lecoanetetal16} found the possibility that mixing induced by convective overshoot was unlikely to disrupt the flame. Consequently, in the absence of a strong magnetic field or substantial mass loss, an off-centered ignition invariably results in the formation of a neutron star \citep{schwabetal16}. 

However, the magnetic field generated through the action of the turbulent MRI and accreted onto the rigidly-rotating HFMWDs will in general be misaligned from its rotational axis. Consequently, the HFMWD will undergo angular momentum loss through the magnetic dipole radiation torque on a characteristic timescale $t_B$,

\begin{equation}
t_B \simeq {I c^3 \over B^2 R^6 \Omega^2}  \approx 10^2 \ {\rm yr}\left(\frac{\beta}{0.1}\right)^{1} \left(\frac{M}{1.4 M_{\odot}}\right) \left(\frac{B}{10^{10} \rm G}\right)^{-2} \left (\Omega_{\rm max} \over \Omega \right)^2 \left(\Omega_{\rm max} \over {5\ {\rm s}^{-1} } \right)^{2/3} .
\label{eqn:tb}
\end{equation}
Here $I$ is the moment of inertia of the WD, $I = \beta M R^2$, where $\beta \simeq 0.1$ \citep{yoonlanger05}, and $\Omega$ is the final angular velocity. 

The subsequent spin evolution of the WD hinges critically upon its mass. A body of literature, beginning with \citet {finnshapiro90} and \citet {shapiroteukolskynakamura90} showed that relativistically-degenerate stars actually {\it spin up} and {\it compress} their cores with angular momentum loss. This seemingly non-intuitive effect is driven fundamentally by the compression induced by angular momentum loss, since the rigidly-rotating near-$M_{\rm Ch}$ WD sequence greatly reduces its moment of inertia as it loses angular momentum. \citet {geroyannispapasotiriou00} expanded on these findings, generating evolutionary sequences of rapidly-rotating magnetic WDs, and demonstrating that near-$M_{\rm Ch}$ WDs with masses $\geq 1.32 M_{\odot}$ continue to spin up, and would compress their cores to extremely high densities $\rho_c = 4 \times 10^{10}$ g cm$^{-3}$, in the absence of nuclear reactions. 

\citet{becerraetal18} showed that WD mergers with accretion rates set by the viscous timescale initially undergo rapid spin up due to the torque exerted by the accreting material. When the Alfvén radius becomes larger than the WD radius, the remnant system enters the propeller phase, spinning down and compressing the WD and causing the central density to rapidly rise. Subsequently, \citet {becerraetal18} demonstrated that the timescale of the evolution of the WD is governed by the magnetic dipole radiation torque timescale $t_B$. In a subsequent paper, \citet {becerraetal19} examined the influence of a strong magnetic field misaligned with the rotational axis of an isolated WD, appropriate for this final phase of evolution of the merger, following the propeller phase. \citet {becerraetal19} demonstrated that the combined effects of neutrino cooling and angular momentum loss due to dipole emission lead to the central nuclear ignition of the merger for magnetic field strengths in excess of $10^7$ G. Their detailed models predict timescales of the same order as our simpler estimate for  $t_B$ \citep {becerraetal19}. \edit1{Moreover, \citet{becerraetal18, becerraetal19} noted the sensitivity of the moment of inertia of the relativistically-degenerate merger to its mass, with small changes of mass responsible for relatively large compression factors. Consequently, we expect that central ignitions of near-$M_{\rm Ch}$ DD mergers should be robust, and will tend to ignite in a very narrow mass range of progenitors subject to compression-induced angular momentum loss.}

As a consequence of the relativistically-degenerate cores of near-$M_{\rm Ch}$ WDs, uniformally-rotating WDs will undergo spin up and compression until they reach carbon ignition in their cores.   The magnetic dipole radiation timescale $t_B$ given by equation \ref{eqn:tb} predicts that for a magnetic field of $10^6$ G, a typical field strength for field WDs \citep{ferrarioetal15}, $t_B$ exceeds the Hubble time. Thus, we emphasize that the production of a HFMWD crucially enables this process of dipole-driven central ignition to be astrophysically relevant. 


\begin{figure}[h]
    \centering
	\begin{center}
	
	\includegraphics[scale=0.35]{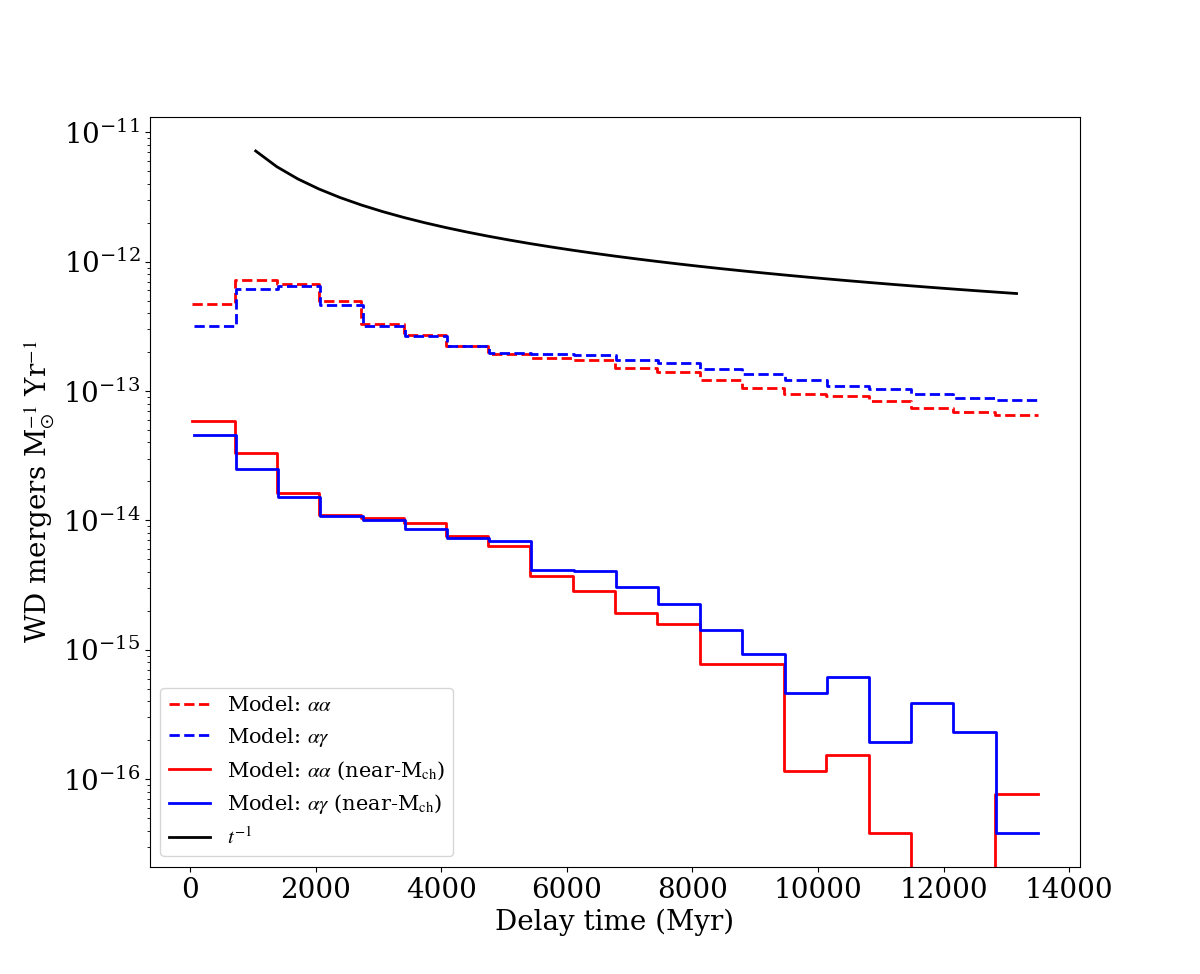}
		\caption{The delay time distribution of double WD mergers, along with the distribution of near-$\rm M_{ch}$ CO mergers, obtained from the $\alpha \alpha$ and $\alpha \gamma$ BPS models. The near-$\rm M_{ch}$ WDs have a total mass $M \geq 1.38 M_{\odot}$ and $M \leq 1.48 M_{\odot}$. }
		\label{fig:dtd_nearMch}
	\end{center}
\end {figure}

\subsection{WD Merger Rate and Delay Time Distribution} \label {sec:mergerrate}

%

Next, to examine the rate of near-$M_{\rm Ch}$ DD mergers and their associated delay time distribution, we employ two  sets of binary population synthesis (BPS) models from \citet{toonenetal2012, toonenetal2017} using the SeBa code \citep{zwart1996,  toonenetal2012}. Both models assume a single burst of star formation, with initial periods determined by a flat distribution in the log of the initial semi-major axis. The models differ in the treatment of the common-envelope (CE) phase during the binary merger \citep {toonenetal2017}. Model $\alpha \alpha$ assumes conservation of orbital energy for every common envelope phase \citep {webbink84}. In contrast, model $\alpha \gamma$ adopts angular momentum conservation \citep {nelemansetal2000}, unless a common envelope phase is initiated by a tidal instability, or if the binary contains a WD. \citet{toonenetal2012, toonenetal2017} provide further details on the BPS models.

Figure \ref{fig:dtd_nearMch} shows the delay time distribution (DTD) of all the WD mergers and the subset of near-$M_{\rm Ch}$ CO WD mergers for both common envelope prescriptions, along with a $\rm t^{-1}$ power law, consistent with the observed delay time distribution \citep{maozetal14}.  To create this DTD plot, we first created 20 equally spaced delay time bins. Then, we divided the number of mergers in each bin by the total BPS simulation mass to obtain the specific number of WD mergers per solar mass. Finally, the specific number of WD mergers is divided by the width of each delay time bin, yielding the specific rate of WD mergers per solar mass per year.

We calculate the fraction of WD mergers producing a double CO WD binary under two assumptions regarding mass accretion and ejection during the merger. In the first model, we assume mass accretion is completely conservative during the merger, without any mass lost through any of the possible processes discussed previously in \S 2 : tidal tails, magnetized jets and outflows, and propeller-driven outflows. In the second model variation, we assume non-conservative mass accretion, with the maximal propelled mass according to eqn. \ref {eqn:mass_ej}. These two model variations span the possible range of mass transfer during the merger. In particular, in order to produce a near-$M_{\rm Ch}$ mass merger, mass ejection requires even more massive, and consequently rarer, WDs. As a result, fully conservative and maximally non-conservative mass accretion mergers yield  upper and lower bounds, respectively, to the fraction of near-$M_{\rm Ch}$ WDs produced. The fraction of near-$M_{\rm Ch}$ CO WD mergers is then estimated by computing the fraction of total post-merger system masses $M$ spanning the range of bounds for rigidly-rotating near $M_{\rm Ch}$ WDs ($1.38 M_{\odot} \leq M \leq 1.48 M_{\odot}$) from the BPS simulation datasets. The $1.38 M_{\odot}$ lower bound mass bound is adopted from the minimum mass of a rigidly-rotating WD which undergoes spin up upon angular momentum loss in full GR, and the upper bound of $1.48 M_{\odot}$ the maximum stable mass of a C-O rigidly-rotating WD  \citep {boshkayevetal13}. 

%


The maximally non-conservative mass transfer assumption yields 2.7\%  and 2.0\%  of near-$M_{\rm Ch}$ CO WD systems  for $\alpha \alpha$ and $\alpha \gamma$ models, respectively. Similarly, under the fully-conservative mass transfer assumption, these fractions increase to 3.5\% and 3.1\% for $\alpha \alpha$ and $\alpha \gamma$ models, respectively. Combined, we estimate that near-$M_{\rm Ch}$ CO WD mergers may account for 2\% - 3.5\% of all WD mergers, with the estimates spanning the range of maximally non-conservative to fully-conservative mass transfer during the merger, and accounting for both the $\alpha \alpha$ and $\alpha \gamma$ common envelope prescriptions.
A radial velocity survey by \citet{maozetal18} shows that the double WD (DWD) merger rate is approximately 6 times the specific SN Ia rate in our galaxy. Thus, near-$M_{\rm Ch}$ DWD mergers may account for up to 12\% - 21\% of all SNe Ia events, if all near-$M_{\rm Ch}$ CO DWD mergers result in a SN Ia.

However, because the fundamental physical process of detonation initiation in a turbulent, unconfined medium is an extremely challenging one to address \citep {poludnenkoetal11, fennplewa17, fisheretal19, poludnenkoetal19}, the explosive outcome of a near-$M_{\rm Ch}$ WD remains uncertain from a first-principles standpoint. A central ignition in the core of the near-$M_{\rm Ch}$ WD leads to a deflagration bubble \citep {nomotoetal84} at a single offset location, according to {\it ab initio} numerical simulations  \citep {zingaleetal11, nonakaetal12, maloneetal14}. This flame bubble will buoyantly rise and become unstable to the reactive Rayleigh-Taylor and Kelvin-Helmholtz instabilities, driving turbulence that may cause the deflagration to transition to a detonation \citep {khokhlov91}, and thereby lead to a SN Ia. Alternatively, the bubble may continue to buoyantly rise until it erupts from the surface of the WD. Under this second possibility, a portion of the buoyant ash will be expelled from the gravitationally-bound WD, which is kicked in the opposite direction and enriched from the remaining ash, which falls back onto the surface of the WD. Observational \citep {jha17} and theoretical evidence \citep {jordanetal12, finketal14} points towards the possibility that at least some of these failed detonation events will be visible as SNe Iax \citep {foleyetal13, stritzingeretal15}, and leave behind surviving kicked WDs enriched by ash rich in IGEs resulting from high-density burning, such as LP 40-365  \citep {raddietal19}.

Therefore, if alternatively a near-$M_{\rm Ch}$ DWD merger ignites but fails to detonate, it will instead produce a SN Iax \citep {jordanetal12, finketal14}. This near-$M_{\rm Ch}$ DWD merger rate is close to the 10 - 30\% of the total SNe Ia rate estimated for SNe Iax by \citet {foleyetal13}, suggesting that near-$M_{\rm Ch}$ DWD mergers may largely account for the SNe Iax population if the majority of mergers ignite as pure deflagrations but fail to detonate. 

\section{Hydrodynamical Simulation and Postprocessing Methodology} \label{sec: methodology}

We next consider the observable signatures of SNe Ia DD near-$M_{\rm Ch}$ events. In order to capture the detailed nucleosynthetic yields and synthetic spectral signatures of SNe Ia events originating \edit1{from} DD near-$M_{\rm Ch}$ mergers, we turn to hydrodynamical simulations, combined with a nucleosynthetic and radiative transfer post-processing pipeline. 


In the following analysis, we will consider 
the possibility that the DD near-$M_{\rm Ch}$ WD progenitor leads to \edit1 {a} successful SNe Ia via the DDT mechanism.
We adopt an equilibrium WD progenitor model with a cold degenerate core and a hot, thick isentropic envelope to model the end state of the merger following spin down and ignition \citep {yoshida_2019}. In particular, as described above, \edit1 {during the final inspiral due to gravitational wave emission, the secondary reaches its Roche limit and is tidally disrupted, forming a massive disk which is then subsequently accreted onto the primary white dwarf.} \edit1 {T}he strong magnetic fields cause nascent HFMWDs to be in a state of uniform rotation. The addition of rigid rotation has only a minimal impact on the density structure of the WD  \citep {ostrikerbodenheimer68, yoshida_2019}, so we utilize a non-rotating WD model for the purpose of computing nucleosynthetic yields and synthetic spectra.  We further neglect the influence of the magnetic field upon the WD structure, as the pressure is dominated by the degenerate electron pressure for HFMWDs with surface fields of $10^{10}$ G and less \citep {jietal13}, and simulate the evolution in pure hydrodynamics.

\edit1 {More specifically, with this physical picture of the merger remnants in mind, we model the hydrostatic initial condition for our time-dependent simulations as follows. The equation of state (EOS) for the hot envelope is the Helmholtz EOS, which includes contributions from nuclei, electrons, and blackbody radiation, and supports an arbitrary degree of degeneracy and special relativity for the electronic contribution \citep{Timmes_2000}, and a completely degenerate EOS for the cold core \citep{shapiroteukolsky86}. The envelope is assumed isentropic. We compute the hydrostatic model using a modified version of Hachisu's self-consistent field method code \citep {hachisu86}, which self-consistently computes both the matter distribution and the gravitational field. The input parameters are the chemical composition of the core and the boundary, the central density, the pressure at which the core-envelope boundary exists, the temperature of the envelope at the core-envelope boundary and the parameter specifying the rotational deformation of the star, which is set zero here. Further details are given in \cite{yoshida_2019}. This hydrostatic model is subsequently mapped into the initial condition for our hydrodynamical simulations, as specified below.}

The central density of a near-$M_{\rm Ch}$ SN Ia event plays a primary role in determining the nucleosynthesis of key iron group elements including $^{56}$Ni as well as neutronized isotopes such as $^{55}$Fe, the parent isotope of $^{55}$Mn.  For illustrative purposes, we adopt only a single set of representative parameters for the merger for this paper. We select a model with a central density $2 \times 10^9$ g cm$^{-3}$, which has been widely adopted for near-$M_{\rm Ch}$ SNe Ia \citep {woosley97, nomotoetal97}. The total mass of the model is $1.378 M_{\odot}$, with a core mass of $0.795 M_{\odot}$ and an envelope mass of $0.593 M_{\odot}$. If the merger process conserved mass, the core and envelope masses would equate to the primary and secondary WD masses, respectively. In reality, mass accretion will be non-conservative due to magnetohydrodynamically-driven jets and winds and tidal tail ejection, implying the secondary mass will be slightly more massive than the $0.593 M_{\odot}$ of the envelope in order to account for mass ejected from the system.

As noted in section \ref  {sec:braking}, the temperature of the envelope is the result of the complex magnetohydrodynamical processes involved during  the merger, including the development of MRI turbulence and a magnetized corona, as well as jets and outflows. In our model, this complexity is subsumed into the entropy of the envelope, or equivalently the temperature at the base of the envelope. The primary WD is modeled as isothermal with a temperature of $10^{8}$  K, while the tidally-disrupted secondary forms a hot, thick  isentropic envelope around the primary with a base temperature $10^{9}$ K\edit1{, typical of binary WD mergers \citep {Zhu_2013}}. The composition of the hydrodynamical model is taken to be 50/50 C/O, though we also consider the effect of solar metallicity in nucleosynthetic post-processing. 


As noted above, while there has been significant advances made in our understanding of turbulent detonation initiation, much work still needs to be done to incorporate these physically-realistic detonation mechanisms into first-principles multidimensional simulations of SNe Ia. However, the {\it ab initio} simulations of the deflagration phase point towards the ignition of a single buoyancy-driven bubble with an mean offset of 40 km \citep {nonakaetal11, zingaleetal11, maloneetal14}. While the process of ignition is inherently a stochastic one, the {\it ab initio} simulations indicate the majority of ignitions lead to buoyancy-driven deflagration phases with weak pre-expansion \citep {fisherjumper15, byrohletal19}. Furthermore, recent advances regarding turbulently-driven detonation initiation point towards a higher DDT transition density, $\sim 10^8$ g cm$^{-3}$, earlier than the distributed burning density of $\sim 10^7$ g cm$^{-3}$ previously assumed \citep {poludnenkoetal19}.

Accordingly, in our DDT model, we adopt a simplified explosion scenario. Incorporating these recent SN IA combustion advances, we neglect the weak pre-expansion of the deflagration phase, and centrally detonate the merger model with a detonation radius of $23$ km.  
The hydrostatic progenitor model is mapped to 2D axisymmetry in the Eulerian adaptive mesh refinement (AMR) code FLASH \citep {Fryxell_2000, Dubey_09, Dubey_14}.\footnote{Though the model presented here is essentially spherically symmetric, the 2D axisymmetric setup enables broader sets of initial conditions for comparison against other published runs.} \edit1{The inviscid Euler equations of hydrodynamics are advanced using an unsplit, higher-order Godunov method. \citep {Lee_2009}} We ran the hydrodynamical simulation on a 2D azimuthally-symmetric cylindrical $r-z$ domain, with $z$ axis extending from $-1.31072 \times 10^6$ km to $1.31072 \times 10^6$ km, and $r$ axis extending from $0$ to $1.31072 \times 10^6$ km.  The region surrounding the WD in the domain is filled with a low density matter with negligible inertia and thermal energy, having an initial density of $10^{-3}$ g cm$^{-3}$ and a temperature of $3 \times 10^{7}$ K. FLASH utilizes \edit1 {the Helmholtz equation of state} \citep {Timmes_2000}. 
An advection-diffusion-reaction equation is employed for the \edit1 {detonation front}, along with a simplified treatment of \edit1 {the nuclear energy generation} \citep {townsleyetal07,townsleyetal09}. 
The Poisson equation for self-gravity is solved using a multipole moment method \citep {Couch_2013} with isolated boundary conditions, including terms up to $l = 6$ in the expansion. In the DDT model, the system is evolved until it reaches free expansion. To follow the nuclear burning regions with maximal refinement, and to derefine regions with low density, we use several refinement criteria based on \citet{townsleyetal07} and \citet{townsleyetal09}. In our run, refinement and derefinement of the grid is based on a dimensionless gradient parameter of density and burnt fraction, a scalar field which tracks the \edit1 {detonation surfaces}, and ranges from 0 for pure fuel to 1 for pure ash \citep{townsleyetal16}. The maximum spatial resolution for the hydrodynamical simulations presented here is $4$ km.

We also incorporate passively-advected Lagrangian tracer particles \citep {dubeyetal12} to record the thermodynamic state along the particle trajectories throughout the simulation. The tracer particles are initialized proportional to the mass in the simulation domain. In the runs presented here, we used $10^4$ particles, which provides good precision for near-$M_{\rm Ch}$ WDs in 2D \citep {seitenzahletal10}. Thermodynamic histories of these particles are restructured into trajectory files which are then post-processed in Torch nuclear network \citep{Timmes_1999}, with 489 species, to obtain the detailed nucleosynthetic yields. A convergence test for sample trajectories showed that the maximum final abundance error for integration time step tolerances using abundance normed errors  between $10^{-6}$ and $10^{-8}$ was $\sim 0.3\%$. Consequently, in the results presented here, an integration timestep tolerance of $10^{-6}$ was used for all tracer particles. 


To generate synthetic spectra,  we map the nucleosynthetic yields onto a 2D axisymmetric velocity mesh. 
This homologously-expanding mesh is then followed with the \edit1 {LTE} radiative transport code SuperNu \citep{wollaeger_2013, wollaeger_2014}, which employs the implicit Monte Carlo (IMC) and discrete diffusion Monte Carlo (DDMC) methods. We bin photons at every time step into 512 logarithmically-spaced wavelength bins within 10$^{3}$ -- 10$^{4.5}$ \AA. To classify the synthetic spectra, we use the supernova identification code SNID \citep{blondin_2011}, which cross correlates the input model spectra with templates of previously-observed events, and identifies the best-matched supernova type as well as its epoch.

\section{Simulation Results} \label{sec: results}


\subsection {DD Near-Chandrasekhar Mass DDT} \label{sec:DDT}

The hydrodynamical simulation of the detonation through free expansion yields a total nuclear energy of $2.13 \times 10^{51} \rm erg$. Overall, the energetics and nucleosynthetic yields are comparable to the most energetic deflagration-to-detonation near-$M_{\rm Ch}$ previously published \citep {seitenzahletal13b, ohlmannetal14, leung_2018}, though with the marked difference of less $^{28} \rm Si$, owing to the less vigorous pre-expansion resulting from our deflagration and turbulently-driven  deflagration-to-detonation model assumptions.  Table \ref{table:1} contains stable mean nucleosynthetic yields for isotopes with abundances greater than $10^{-6} M_{\odot}$. Complete mean nucleosynthetic yields, which include both stable and radioisotopes obtained at the onset of free expansion, $t = 2.5$ s, with abundances greater than $10^{-6} M_{\odot}$ are given in Table \ref{table:2}.

\begin{deluxetable}{ccc}
\tabletypesize{\footnotesize}
\tablecolumns{10}
\tablewidth{0pt}
\tablehead{
\colhead{Isotope} & \colhead{$Z = 0$} & \colhead{$Z = Z_{\odot}$}}
\startdata
$^{4} \rm He$ & $9.21 \times 10^{-3}$ & $8.09 \times 10^{-3}$ \\
$^{12} \rm C$ & $2.85 \times 10^{-6}$ & $3.53 \times 10^{-6}$ \\
$^{16} \rm O$ & $9.33 \times 10^{-4}$ & $1.04 \times 10^{-3}$ \\
$^{24} \rm Mg$ & $7.78 \times 10^{-5}$ & $1.55 \times 10^{-5}$ \\
$^{28} \rm Si$ & $7.76 \times 10^{-3}$ & $7.95 \times 10^{-3}$\\
$^{29} \rm Si$ & $1.97 \times 10^{-6}$ & $3.84 \times 10^{-6}$ \\
$^{30} \rm Si$ & $8.23 \times 10^{-6}$ & $4.68 \times 10^{-6}$ \\
$^{31} \rm P$ & $1.81 \times 10^{-6}$ & $3.47 \times 10^{-6}$ \\
$^{32} \rm S$ & $4.93 \times 10^{-3}$ & $4.80 \times 10^{-3}$ \\
$^{33} \rm S$ & $2.59 \times 10^{-7}$ & $3.62 \times 10^{-6}$ \\
$^{34} \rm S$ & $3.93 \times 10^{-6}$ & $2.64 \times 10^{-5}$ \\
$^{35} \rm Cl$ & $1.43 \times 10^{-6}$ & $2.96 \times 10^{-6}$ \\
$^{36} \rm Ar$ & $1.22 \times 10^{-3}$ & $1.12 \times 10^{-3}$ \\
$^{38} \rm Ar$ & $5.69 \times 10^{-7}$ &$1.60 \times 10^{-5}$ \\
$^{39} \rm K$ & $1.70 \times 10^{-6}$ & $4.23 \times 10^{-6}$ \\
$^{40} \rm Ca$ & $1.32 \times 10^{-3}$ & $1.18 \times 10^{-3}$ \\
$^{44} \rm Ca$ & $1.10 \times 10^{-5}$ & $8.67 \times 10^{-6}$ \\
$^{48} \rm Ti$ & $6.29 \times 10^{-5}$ & $5.55 \times 10^{-5}$ \\
$^{49} \rm Ti$ & $5.09 \times 10^{-7}$ & $2.67 \times 10^{-6}$ \\
$^{51} \rm V$ & $1.12 \times 10^{-5}$ & $1.59 \times 10^{-5}$ \\
$^{50} \rm Cr$ & $3.88 \times 10^{-5}$ & $4.74 \times 10^{-5}$ \\
$^{52} \rm Cr$ & $1.26 \times 10^{-3}$ & $1.19 \times 10^{-3}$ \\
$^{53} \rm Cr$ & $2.31 \times 10^{-4}$ & $3.07 \times 10^{-4}$ \\
$^{55} \rm Mn$ & $9.19 \times 10^{-3}$ & $1.02 \times 10^{-2}$ \\
$^{54} \rm Fe$ & $7.79 \times 10^{-2}$ & $8.52 \times 10^{-2}$ \\
$^{56} \rm Fe$ & $9.98 \times 10^{-1}$ & $9.60 \times 10^{-1}$ \\
$^{57} \rm Fe$ & $4.24 \times 10^{-2}$ & $4.72 \times 10^{-2}$ \\
$^{59} \rm Co$ & $1.45 \times 10^{-3}$ & $1.54 \times 10^{-3}$ \\
$^{58} \rm Ni$ & $1.98 \times 10^{-1}$ & $2.26 \times 10^{-1}$ \\
$^{60} \rm Ni$ & $1.78 \times 10^{-2}$ & $1.45 \times 10^{-2}$ \\
$^{61} \rm Ni$ & $7.18 \times 10^{-4}$ & $7.90 \times 10^{-4}$ \\
$^{62} \rm Ni$ & $4.61 \times 10^{-3}$ & $6.42 \times 10^{-3}$ \\
$^{63} \rm Cu$ & $4.75 \times 10^{-5}$ & $6.58 \times 10^{-6}$ \\
$^{65} \rm Cu$ & $5.22 \times 10^{-6}$ & $5.69 \times 10^{-6}$ \\
$^{64} \rm Zn$ & $2.77 \times 10^{-4}$ & $8.66 \times 10^{-5}$ \\
$^{66} \rm Zn$ & $8.67 \times 10^{-5}$ & $1.17 \times 10^{-4}$ \\
$^{68} \rm Zn$ & $2.04 \times 10^{-6}$ & $1.35 \times 10^{-7}$ \\
$^{70} \rm Ge$ & $2.31 \times 10^{-6}$ & $2.31 \times 10^{-6}$ \\
\enddata
\caption{DDT stable mean nucleosynthetic yields in solar masses. Only isotopes in excess of $10^{-6}\ M_{\odot}$ in either model are shown.} 
\tablecomments{Abundances for all isotopes are provided in machine-readable format.}
\label{table:1}
\end{deluxetable}

\begin{deluxetable}{ccc}
\tabletypesize{\footnotesize}
\tablecolumns{10}
\tablewidth{0pt}
\tablehead{
\colhead{Isotope} & \colhead{$Z = 0$} & \colhead{$Z = Z_{\odot}$}}
\startdata
$^{4} \rm He$ & $9.21 \times 10^{-3}$ & $8.09 \times 10^{-3}$ \\
$^{12} \rm C$ & $2.85 \times 10^{-6}$ & $3.53 \times 10^{-6}$ \\
$^{16} \rm O$ & $9.33 \times 10^{-4}$ & $1.04 \times 10^{-3}$ \\
$^{24} \rm Mg$ & $7.78 \times 10^{-5}$ & $1.55 \times 10^{-5}$ \\
$^{28} \rm Si$ & $7.76 \times 10^{-3}$ & $7.95 \times 10^{-3}$ \\
$^{29} \rm Si$ & $7.47 \times 10^{-7}$ & $3.84 \times 10^{-6}$ \\
$^{30} \rm Si$ & $1.55 \times 10^{-7}$ & $4.65 \times 10^{-6}$ \\
$^{30} \rm P$ & $5.85 \times 10^{-6}$ & $3.38 \times 10^{-8}$\\
$^{30} \rm S$ & $2.22 \times 10^{-6}$ & $7.15 \times 10^{-12}$ \\
$^{31} \rm S$ & $9.57 \times 10^{-7}$ &  $5.44 \times 10^{-8}$ \\
$^{32} \rm S$ & $4.93 \times 10^{-3}$ & $4.80 \times 10^{-3}$ \\
$^{34} \rm S$ & $1.62 \times 10^{-6}$ & $2.64 \times 10^{-5}$ \\
$^{34} \rm Cl$ & $1.65 \times 10^{-6}$ & $5.59 \times 10^{-9}$ \\
$^{35} \rm Cl$ & $1.11 \times 10^{-6}$ & $2.94 \times 10^{-6}$ \\
$^{36} \rm Ar$ & $1.22 \times 10^{-3}$ & $1.12 \times 10^{-3}$ \\
$^{38} \rm Ar$ & $1.75 \times 10^{-7}$ & $1.60 \times 10^{-5}$ \\
$^{39} \rm K$ & $1.61 \times 10^{-6}$ & $4.17 \times 10^{-6}$ \\
$^{40} \rm Ca$ & $1.32 \times 10^{-3}$ & $1.18 \times 10^{-3}$ \\
$^{44} \rm Ti$ & $1.10 \times 10^{-5}$ & $8.67 \times 10^{-6}$ \\
$^{48} \rm Cr$ & $6.29 \times 10^{-5}$ & $5.55 \times 10^{-5}$ \\
$^{49} \rm Cr$ & $5.08 \times 10^{-7}$ & $2.67 \times 10^{-6}$ \\
$^{50} \rm Cr$ & $3.88 \times 10^{-5}$ & $4.74 \times 10^{-5}$ \\
$^{51} \rm Mn$ & $1.11 \times 10^{-5}$ & $1.58 \times 10^{-5}$ \\
$^{53} \rm Mn$ & $3.25 \times 10^{-6}$ & $4.29 \times 10^{-6}$ \\
$^{52} \rm Fe$ & $1.26 \times 10^{-3}$ & $1.19 \times 10^{-3}$ \\
$^{53} \rm Fe$ & $2.28 \times 10^{-4}$ & $3.03 \times 10^{-4}$ \\
$^{54} \rm Fe$ & $7.79 \times 10^{-2}$ & $8.52 \times 10^{-2}$ \\
$^{55} \rm Fe$ & $1.55 \times 10^{-4}$ & $1.83 \times 10^{-4}$ \\
$^{56} \rm Fe$ & $5.38 \times 10^{-5}$ & $7.85 \times 10^{-5}$ \\
$^{55} \rm Co$ & $9.04 \times 10^{-3}$ & $1.00 \times 10^{-2}$ \\
$^{56} \rm Co$ & $1.48 \times 10^{-4}$ & $1.60 \times 10^{-4}$ \\
$^{57} \rm Co$ & $8.77 \times 10^{-5}$ & $1.04 \times 10^{-4}$ \\
$^{56} \rm Ni$ & $9.98 \times 10^{-1}$ & $9.60 \times 10^{-1}$ \\
$^{57} \rm Ni$ & $4.24 \times 10^{-2}$ & $4.71 \times 10^{-2}$ \\
$^{58} \rm Ni$ & $1.98 \times 10^{-1}$ & $2.26 \times 10^{-1}$ \\
$^{59} \rm Ni$ & $1.98 \times 10^{-4}$ & $2.21 \times 10^{-4}$ \\
$^{60} \rm Ni$ & $1.35 \times 10^{-4}$ & $1.63 \times 10^{-4}$ \\
$^{58} \rm Cu$ & $2.80 \times 10^{-4}$ & $2.22 \times 10^{-6}$ \\
$^{60} \rm Cu$ & $8.22 \times 10^{-5}$ & $6.92 \times 10^{-5}$ \\
$^{61} \rm Cu$ & $1.18 \times 10^{-5}$ & $1.33 \times 10^{-5}$ \\
$^{60} \rm Zn$ & $1.76 \times 10^{-2}$ & $1.43 \times 10^{-2}$ \\
$^{61} \rm Zn$ & $7.07 \times 10^{-4}$ & $7.77 \times 10^{-4}$ \\
$^{62} \rm Zn$ & $4.61 \times 10^{-3}$ & $6.42 \times 10^{-3}$ \\
$^{63} \rm Zn$ & $4.38 \times 10^{-6}$ & $3.29 \times 10^{-6}$ \\
$^{63} \rm Ga$ & $4.26 \times 10^{-5}$ & $3.28 \times 10^{-6}$ \\
$^{64} \rm Ga$ & $4.65 \times 10^{-6}$ & $1.93 \times 10^{-6}$ \\
$^{64} \rm Ge$ & $2.72 \times 10^{-4}$ & $8.46 \times 10^{-5}$ \\
$^{65} \rm Ge$ & $4.69 \times 10^{-6}$ & $5.09 \times 10^{-6}$ \\
$^{66} \rm Ge$ & $8.66 \times 10^{-5}$ & $1.17 \times 10^{-4}$ \\
$^{68} \rm Se$ & $1.92 \times 10^{-6}$ & $8.50 \times 10^{-8}$\\
$^{70} \rm Se$ & $2.31 \times 10^{-6}$ & $2.84 \times 10^{-6}$\\
\enddata
\caption{DDT frozen-out mean nucleosynthetic yields in solar masses. Only isotopes in excess of $10^{-6}\ M_{\odot}$ in either model are shown.} 
\tablecomments{Abundances for all isotopes are provided in machine-readable format.}
\label{table:2}
\end{deluxetable}

\begin{figure}[h]
    \centering
	\begin{center}
	
	\includegraphics[scale=0.20]{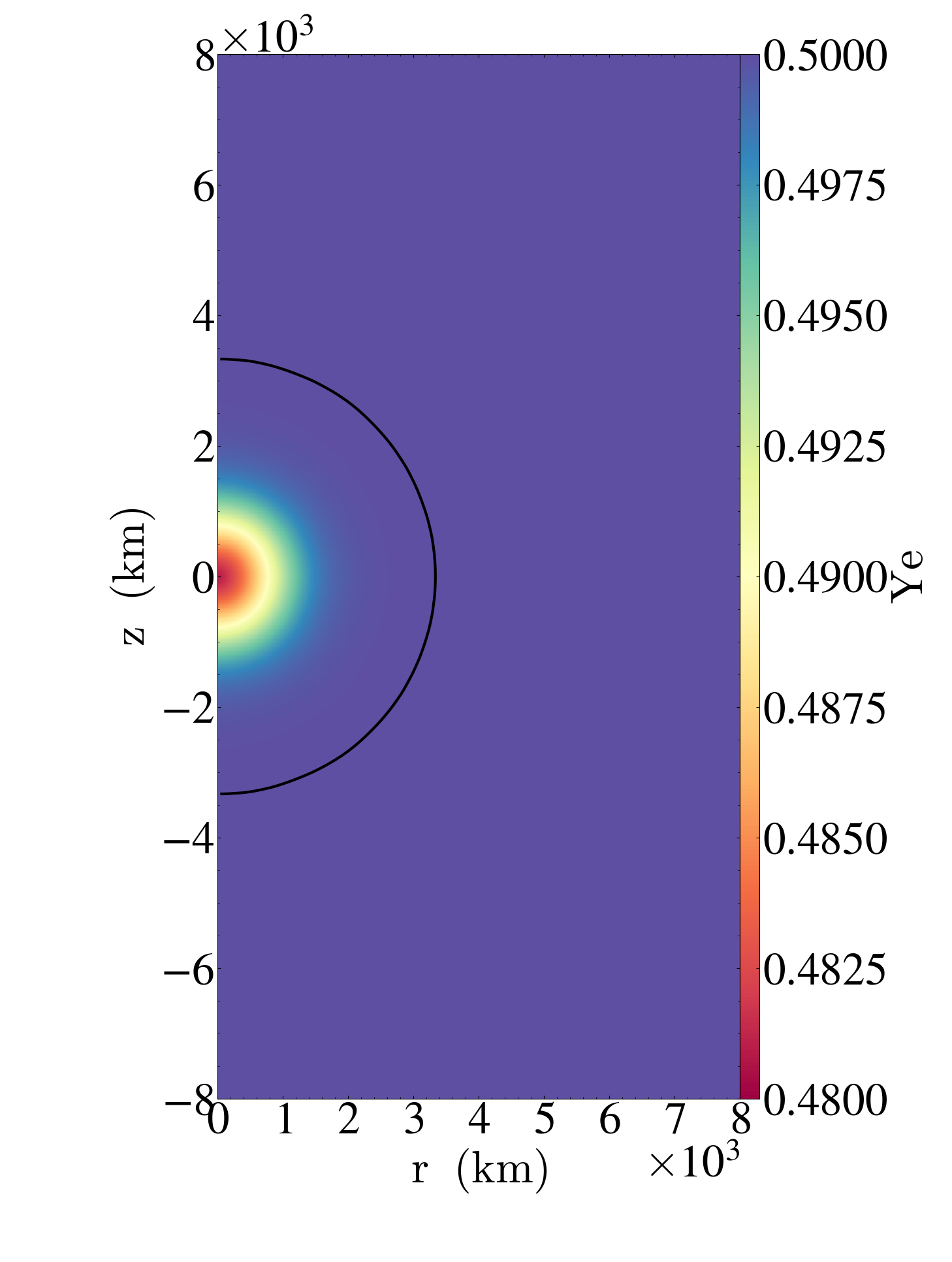}
		\caption{Slice plot of $Y_e$ at $t=$ 240 ms, when the nucleosynthetic yields have frozen out, showing the centrally-concentrated region of neutronized isotopes produced from electron captures. Note that this plot represents the hydrodynamic model output, which is at zero WD progenitor metallicity. The $10^6$ g cm$^{-3}$ density isocontour is represented by a black line.}
	    \label{fig:ye_slice}
	\end{center}
\end {figure}

Plots of the frozen-out electron abundance fraction $Y_e$ = $\sum_i X_i Z_i/A_i$, where $Z_i$ is the atomic number, $A_i$ is the atomic mass number, and $X_i$ is the mass abundance of the isotope $i$, are depicted near the onset of homologous expansion in Figure \ref{fig:ye_slice}. Consistent with the findings of previous simulations of near-$M_{\rm Ch}$ SNe Ia DDT models (including \citealt {meyeretal96, nomotoetal97, woosley97, brachwitzetal00, kruegeretal12}), we find that the greater electron capture rates in the cores of denser WD progenitors lead to greater abundances of stable IGEs in the n-NSE regime. 

Next, in conducting our synthetic spectroscopic classification, we compare not to a single event, but rather to a template library, just as new optical transients are classified observationally. Accordingly, we utilize the supernova identification code SNID, which defines the quality parameter rlap to quantify the strength of the correlation between the input synthetic spectra and previously observed events  \citep{blondin_2011}. \cite{tonry_1979} defined the parameter $r$ in terms of the ratio of the height $h$ to the root mean square of the antisymmetric term of the normalized cross-correlation function between the input and the template spectra, $\sigma_{a}$, so that $r = 2^{-1/2} h \sigma_{a}^{-1}$.
 In addition, SNID also uses the overlap of the spectra in log wavelength 
 space to quantify the reliability of the correlation peak. To match with the wavelength range of the input spectrum at the correlation peak redshift, the template spectra are trimmed. Then the overlap parameter, named lap, in log wavelength space 
is given by $0\leq$ lap $\leq \ln (\lambda_{1} / \lambda_{0} )$, where $[\lambda_{0}, \lambda_{1}]$ is the rest-frame wavelength range of the input spectrum. Values of rlap, defined as the product of $r$ and lap, over 5 are considered to be good matches. We constrain our synthetic spectra to the rest frame at redshift $z = 0$.

\begin{figure}[ht!]
\centering
   \includegraphics[width=0.7\textwidth]{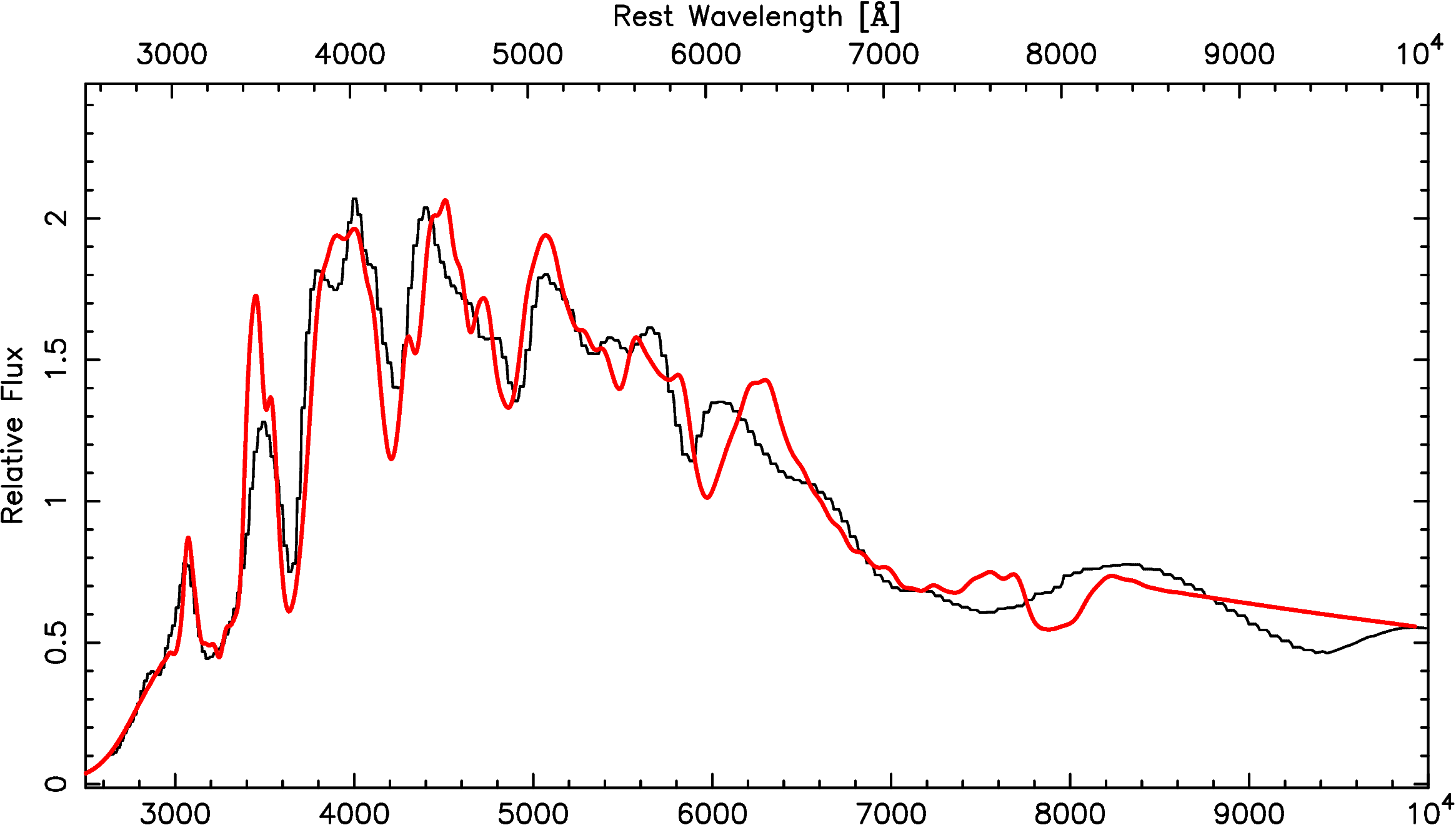}
   \caption{Synthetic peak brightness model spectrum (black) at WD progenitor metallicity $Z = Z_{\odot}$ compared with the best matching spectral template, SN 1999ee, a normal Ia (red).}
   \label{fig:lowden_spectra} 
\end{figure}

SNID classifies the best-matches to the synthetic spectra at peak-brightness for both WD progenitor metallicities $Z = 0$ and $Z = Z_{\odot}$ as SN 1999ee, a type Ia-norm, at an epoch of -9.1 d \citep{hamuy2002_99ee}. These matches have rlap scores 7.5 and 8.6, respectively. Figure \ref{fig:lowden_spectra} shows the synthetic spectrum for $Z = Z_{\odot}$, plotted against its best-match spectrum. {\it We emphasize that the match is obtained directly from the near-$M_{\rm Ch}$ progenitor without any fine-tuned parameters.} The overall agreement between the two spectra is quite good, with the noted exception of a blueshift in the synthetic spectral Si II absorption feature near 6000 {\AA}, in comparison to SN 1999ee, as well as in the pre-peak epoch identification. The inclusion of non-LTE in the radiative transfer is expected to shift the Si II feature towards the red, which may partially account for this disagreement \citep {shenetal21}. The blueward shift of the Si II feature caused by the LTE approximation may also partially account for the SNID epoch identification prior to peak brightness, as the Si II feature generally shifts redward at later epochs \citep {pereiraetal13}. 

We also note that SN 1999ee is a member of a class of SNe Ia with high-velocity Si II and Ca II features (HVFs). The origin of these features in SN 1999ee and other similar SNe Ia HVF events has been discussed in the literature, with possibilities ranging from density and abundance enhancements in the progenitor WD to interactions with circumstellar material \citep {mazzalietal05}. Our models support the viewpoint that the high velocity Si II HVF in particular may be connected to nuclear burning on massive white dwarfs \citep {katoetal18}.

To compare against observations of supernova remnants, we consider stable isotopic abundances. Because supernova remnants typically have ages of hundreds to thousands of years, in the decay chain $^{53}_{26} \rm Fe$ $\xrightarrow{\text{9m}}$ $^{53}_{25} \rm Mn$ $\xrightarrow{\text{4Myr}}$ $^{53}_{24} \rm Cr$, we restrict the $^{53}_{26} \rm Fe$ decay to $^{53}_{25} \rm Mn$ only \citep{unterweger92}.  The decayed IGE abundance ratios relative to solar abundances, with a WD progenitor with solar metallicity $Z_{\odot} = 0.0139$ \citep{asplund2021}, and  C/O ratio of unity yield [Ni/Fe] = 0.62, [Cr/Fe] = -1.11 and [Mn/Fe] = 0.02, all relative to solar. The [Ni/Fe] ratio is higher than Tycho or Kepler, but comparable to SNR 3C 397. We emphasize that this comparison is based upon our specific choice of parameters, and is therefore illustrative rather than exhaustive of the range of possibilities of DD near-$M_{\rm Ch}$ SNe Ia.



\section{Observational Implications} 
\label{sec:implications}

The classical single-degenerate near-$M_{\rm Ch}$  DDT mechanism requires fine-tuning of the ignition in order to obtain the vigorous pre-expansion necessary for a normal SN Ia. In contrast, DD near-$M_{\rm Ch}$ SNe Ia naturally begin with an extended heated envelope of tidally-disrupted material from the secondary. The braking timescale is less than the cooling timescale for the WD merger for a HFMWD, allowing the WD progenitor to retain this extended envelope at the onset of detonation. As a direct result, a DDT within a DD near-$M_{\rm Ch}$ naturally predicts a normal SN Ia with high velocity Si II features without any fine-tuning of parameters, even though the silicon abundance is significantly sub-solar. \edit1{We expect that a range of central densities at ignition, and entropies of the envelopes of DD near-$M_{\rm Ch}$ progenitors will be produced from DD near-$M_{\rm Ch}$ SNe Ia and SNe Iax, depending on the masses of the binary WDs. These progenitors will in turn produce a range of nucleosynthesized $^{56}$Ni and hence intrinsic luminosities.  Since the compression mechanism of  DD near-$M_{\rm Ch}$ due to angular-momentum loss differs fundamentally from the compression due to accretion from the non-degenerate donor of the SD scenario, further models are required to explore the evolution of highly-magnetized WD mergers as they reach ignition conditions, and connect these to SNe Ia and SNe Iax explosion models.}

A similar magnetic dipole mechanism for nuclear ignition as discussed here is also envisioned for the core degenerate scenario \citep {ilkovsoker12}. The two scenarios differ in the physical explanation for the delay-time distribution. In the core degenerate scenario, the delay-time distribution is explained by the magnetic braking timescale, whereas in our scenario the DTD is explained instead by gravitational wave inspiral, as in the standard double-degenerate scenario. Because the core degenerate scenario predicts a long post-merger delay, and our proposed scenario a relatively prompt delay, the two scenarios offer different possibilities for SN ejecta interactions, which can be used in observational tests.

SNe Ia and their remnants are known to have a high degree of spherical symmetry. Spectropolarimetry of normal SNe Ia generally reveal a low degree of polarization, typically 0.5\% in the Si II absorption line feature at peak brightness, which is a sensitive probe of the outskirts of the SN  \citep {wangwheeler08, cikotaetal19}. This degree of polarization in SNe Ia is particularly striking in comparison with core-collapse SNe, particularly stripped envelope CCSNe, which can exhibit much greater polarizations up to several percent, which translate into intrinsic SN asymmetries of tens of percent \citep {wangetal01, wangwheeler08}. Additionally, the line and thermal X-ray emission of SNe Ia remnants also possess a high degree of spherical symmetry \citep {lopezetal09, lopezetal11}. Taken together, these observational findings place stringent limits on SNe Ia channels which predict large asymmetries, either in the progenitor or the explosion mechanism, or through the SN blast wave interacting with a non-degenerate companion or circumstellar medium. For example, the violent double degenerate merger predicts Si II polarization of $\sim 1 - 3\%$, in excess of most observations of normal SNe Ia  \citep {cikotaetal19}. Additionally, the interaction of the SN blast wave with the non-degenerate stellar companion in the classical single-degenerate near-$M_{\rm Ch}$ scenario leaves a large ``shadow'' which is present in the young remnant for hundreds of years \citep {grayetal16}.

In contrast, the DD near-$M_{\rm Ch}$ channel predicts the WD merger progenitor will be uniformly and slowly rotating at the time of detonation, and so will be very highly spherical. Furthermore, as emphasized in our discussion of  deflagration initiation and the turbulent deflagration-to-detonation mechanism, recent work points towards a mild pre-expansion and relatively prompt detonation. Consequently, DD near-$M_{\rm Ch}$ SNe Ia will have very low Si II polarization near peak brightness.\footnote{The failed detonation of  DD near-$M_{\rm Ch}$ progenitors leading to SNe Iax will have a much greater degree of assymmetry originating from the explosion mechanism itself: the pure deflagration within its interior.} While the first DDT simulations of  DD near-$M_{\rm Ch}$ SNe Ia presented in this paper neglect the initial deflagration phase, we expect that more realistic simulations including the initial deflagration phase through the DDT will produce Si II polarization levels similar to prior near-$M_{\rm Ch}$ SNe Ia DDT models, typically well under 1\%, in line with observational constraints \citep {cikotaetal19}. Further, DD near-$M_{\rm Ch}$ SNe Ia remnants will also naturally be nearly spherically symmetric, without the shadow predicted to arise from the companion in the classical single-degenerate near-$M_{\rm Ch}$ scenario. 

A large body of observational evidence has also placed stringent constraints on SNe Ia nebular H $\alpha$ \citep {leonard07, tuckeretal20}, X-ray \citep {marguttietal14}, and radio emission \citep {chomiuketal16}, and the existence of stellar binary companions \citep {lietal11} and ex-companions \citep {schaferpagnotta12} pre-explosion and post-explosion, respectively, in the vast majority of normal SNe Ia systems. A natural prediction of the near-$M_{\rm Ch}$ DD scenario is that there is no surviving companion, implying that there can also be no nebular H $\alpha$ emission. The near-$M_{\rm Ch}$ DD scenario further predicts an absence of strong prompt X-ray and radio emission, although there are prospects for delayed interaction with the propelled and ejected material. 

Recent work on the ratio on the abundances of Ni II / Fe II, accessible through optical and NIR lines, has elucidated that this line ratio is sensitive to the mass of the progenitor, owing largely to the production of stable Ni isotopes in the cores of near-$M_{Ch}$ events, particularly $^{58}$Ni \citep {florsetal18, florsetal20}. The use of stable Ni is a therefore a promising discriminant between near-$M_{\rm Ch}$ and sub-$M_{\rm Ch}$ events  (though also subject to caveats -- see  \citep {blondinetal21}). The nucleosynthetic yields presented here predict a relatively high abundance of $^{58}$Ni / $^{56}$Fe = 0.2 for a DD near-$M_{\rm Ch}$ event, in comparison to other previously-computed near-$M_{\rm Ch}$ models \citep {blondinetal21}. While the  $^{58}$Ni / $^{56}$Fe abundance ratio will decrease in a more realistic explosion scenario including the deflagration phase, the stable Ni test may help shed light not only upon the mass of the SN Ia progenitor, but also the physics of its formation and the detonation mechanism. 

While substantial evidence points towards the likelihood that partially burnt near-$M_{\rm Ch}$ WDs lead to SNe Iax \citep [eg,][] {jha17}, there is a relative scarcity of direct observational evidence for helium through the helium-accreting single degenerate channel. \citet {jacobsongalan19} find early-time He I CSM lines in two of a sample of 44 SNe Iax; with a luminosity limit cutoff, they conclude He I CSM features are present in 33\% of the remaining sample. They further find no evidence for stripped helium in the late-time nebular spectra of 11 SNe Iax, placing an upper bound limit on the amount of helium to be less than $10^{-2} M_{\odot}$ in these events, with even tighter bounds of $10^{-3} M_{\odot}$ for stripped hydrogen.  In the context of the near-$M_{\rm Ch}$ DD scenario, the absence of large amounts of helium in SNe Iax spectra is a natural consequence of the complete merger of the double degenerate system prior to the SN Iax event. Most of the thin $10^{-3} - 10^{-1} M_{\odot}$ \citep {lawlormacdonald06} helium surface layers atop the two white dwarfs will be susceptible to nuclear burning during the final merger. However, the rapid accretion just prior to and during the tidal disruption of the donor may cause a small amount of helium to be ejected through the outer Lagrange point L2 \citep {livioetal79, macleodetal18}. Such ejected helium may account for the early-time He I CSM seen in some SNe Iax.


Another strong constraint on SNe Ia channels is the $t^{-1}$ observed delay time distribution (DTD) \citep {maozetal10, maozbadenes10, grauretal11, maozetal12}. DD models generically predict a $t^{-1}$ DTD from gravitational wave inspiral. In contrast, theoretical models of the SD typically produce a sharper decay owing the absence of the requisite intermediate mass SD donors over 10 Gyr timescales. While multi-component SD DTDs invoking both main sequence and red giant donor channels have found agreement with a $t^{-1}$ DTD \citep {hachisuetal08}, these same models encounter the aforementioned tensions with the observed absence of companion signatures in the majority of SNe Ia events. In contrast, DD near-$M_{\rm Ch}$ SNe Ia originate from the same gravitational wave inspiral mechanism which underlies the overall DD population and do not possess companion signatures. Consequently, as we have demonstrated in section \ref {sec:mergerrate}, the DTD of DD near-$M_{\rm Ch}$ SNe Ia retains the $t^{-1}$ DTD of the wider DD population. Furthermore, the characteristic $t^{-1}$ DTD shape is not significantly biased by a mass selection effect, even taking into account the maximum possible mass lost through the propeller phase.

Recent models of double-detonations on sub-$M_{\rm Ch}$ WDs suggest that carbon-enriched helium surface layers may produce sufficient  $^{55}$Mn to reduce but not entirely eliminate the need for near-$M_{\rm Ch}$ SNe Ia events  \citep{gronowetal21}. Consequently, the classical SD scenario and its corresponding predictions for companion and ex-companion interactions is in tension with the requisite need to produce $^{55}$Mn through near-$M_{\rm Ch}$. The nucleosynthetic yields for the DD near-$M_{\rm Ch}$ event presented here shows an approximately solar abundance of Mn/Fe. Late-time light curves \citep {seitenzahletal09} offer a window into the 55 and 57 isobar decay chains terminating in $^{55}$Mn and $^{57}$Fe, respectively, and may help shed light on both the progenitor as well as the explosion scenario.

There are two primary means of observationally determining the galactic WD merger rate. The first method infers the WD merger rate from observations of WD field binaries \citep {maozetal18}. The second examines single WDs, and discriminates those produced by mergers from those resulting from isolated evolution \citep {chengetal19}. The field WD merger rate obtained in the most recent observations is systematically higher than that inferred from merged single WDs, which may be at least in part be due to unaccounted systematics between these two techniques. Explosive transients, including the near-$M_{\rm Ch}$ DD mergers described here, may account for a key systematic bias between these two sets of observations. 

Observational surveys of the abundance ratios of [Ni/Fe] and [Mn/Fe] have yielded a rich picture of the nucleosynthetic histories of different galactic environments. In particular, the dwarf spheroidal Sculptor, which had a single burst of star formation after its formation,  has decreasing ratios of [Ni/Fe] and [Mn/Fe] as a function of metallicity [Fe/H], a proxy for age \citep {kirbyetal19, delosreyesetal20}. In contrast, other galactic environments with ongoing star formation histories, including Leo I, Fornax, and the Milky Way, exhibit increasing ratios of [Ni/Fe] and [Mn/Fe] with [Fe/H] \citep {kirbyetal19, delosreyesetal20}. Because near-$M_{\rm Ch}$ events, including the DD channel presented here, generally have higher  [Ni/Fe] and [Mn/Fe] in comparison to sub-$M_{\rm Ch}$ events, the nucleosynthetic histories of stellar environments such as Sculptor undergoing a single star formation burst point towards an increasing predominance of sub-$M_{\rm Ch}$ relative to near-$M_{\rm Ch}$ events with time. Other stellar environments with ongoing star formation histories point instead towards an increasing ratio of  near-$M_{\rm Ch}$ to sub-$M_{\rm Ch}$  events with time. A possible explanation for these diverse SNe Ia histories stems from considerations of both the  sub-$M_{\rm Ch}$ as well as the DD near-$M_{\rm Ch}$ DTDs. The DTD of sub-$M_{\rm Ch}$ double detonations from \edit1{non-degenerate donors inferred from BPS models exhibit a sharp cutoff at times earlier than $\sim 1$ Gyr \citep {ruiteretal14}.}
In contrast, the characteristic $t^{-1}$ shape of the DD near-$M_{\rm Ch}$ DTD \edit1{is significantly delayed relative to sub-$M_{\rm Ch}$ double detonations fed by non-degenerate donors.}
\edit1 {Furthermore,} the qualitative difference of the  DD near-$M_{\rm Ch}$ DTD  with the classical SD near-$M_{\rm Ch}$ DTD, which cuts off sharply after a few Gyr  \citep {kobayashietal19}, may offer a potentially observable test of these different scenarios for the production of  near-$M_{\rm Ch}$ SNe Ia.

During the final writing of this paper, the discovery of a rapidly-rotating, strongly-magnetized near-$M_{\rm Ch}$ WD, ZTF J190132.9+145808.7, was reported \citep {caiazzo2021}. The surface magnetic field was estimated to be 600 - 800 MG, with an inferred internal O/Ne composition approaching a central density of $\sim$ several $\times 10^9$ g cm$^{-3}$ and central temperature $\sim 7 \times 10^8$ K. This rapidly-rotating, strongly-magnetized near-$M_{\rm Ch}$ WD is similar to the DD near-$M_{\rm Ch}$ progenitors discussed in this paper. The inferred internal composition and thermodynamic structure, as well as the inferred age of this WD will, however, hinge crucially on magnetohydrodynamical processes which give rise to such a strongly-magnetized WD, and further modeling is required \edit1{to better elucidate its formation mechanism and evolution.} 


     

\section{Conclusion}

In this paper, we have presented a model for the production of near-$M_{\rm Ch}$ WDs from the DD channel resulting from the merger and inspiral of two C/O WDs. The magnetic field plays a crucial role in this mechanism, first during the production of a high field magnetic WD during the merger, and secondly in compressing the core of the WD to ignition conditions during spin down. We have also demonstrated the onset of a propeller phase during the merger, with potentially observable interaction signatures. \edit2{Finally, we presented an exploratory hydrodynamical simulation of a DD near-$M_{\rm Ch}$ system, computed its nucleosynthetic yields, and matched its synthetic spectrum to that of a normal SN Ia. The initial condition of this hydrodynamical simulation incorporated the effect of tidal heating of the secondary WD during the merger, using a self-consistent equilibrium with an isentropic hot envelope. The hydrodynamical simulation assumes negligible pre-expansion and negligible magnetic pressure and rotational support within the merger.}

To summarize the key observable signatures of SNe Ia and SNe Iax produced by DD near-$M_{\rm Ch}$:

\begin {itemize}

 \item A $t^{-1}$ DTD both DD near-$M_{\rm Ch}$ SNe and SNe Iax.
 
 \item The absence of companion and ex-companion signatures in both DD near-$M_{\rm Ch}$ SNe and SNe Iax.
 
 \item Weakly-polarized Si II lines in DD near-$M_{\rm Ch}$ SNe Ia.
 
 \item In  near-$M_{\rm Ch}$ DD SNe Ia, the production of a bright normal to potentially overluminous (91T-like) event. 
 
  \item The absence of strong H and He signatures  in both DD near-$M_{\rm Ch}$ SNe and SNe Iax. Trace He abundaces may be possible by ejection through the outer Lagrange point L2, and may account for He I CSM in some events.
  
 \item Solar or super-solar abundances of [Mn/Fe] and [Ni/Fe].

 \item \edit1 {Highly-spherical SNRs, with high [Mn/Fe] and [Ni/Fe] abundance ratios, and without companion shadows.}
 
\end {itemize}

Further theoretical modeling and simulation is required to obtain a clearer picture of the role which magnetohydrodynamics plays in WD mergers, and help address which circumstances lead to explosive transients, and which lead to stable massive magnetized mergers. While magnetohydrodynamics plays a key role during the WD merger, the process has been long neglected in the majority of models. Massive magnetized WDs like  ZTF J190132.9+145808.7 are providing us with a strong clue that the stellar merger, magnetohydrodynamics, and the SN Ia mechanism are all inextricably linked.

New observational instruments, including JWST and \edit1 {the Rubin Observatory}, will help discriminate both between sub-$M_{\rm Ch}$ and super-$M_{\rm Ch}$ events and the explosion scenario using a wide range of observational tests, including the radioisotopic decay of SNe Ia late-time light curves, and the presence of stable Ni in the nebular phase of SNe Ia. Detailed surveys of galactic SN Ia SNRs, particularly of the near-$M_{\rm Ch}$ SNR 3C 397,  will shed a complementary light on the stellar progenitor and the explosion mechanism. 

\acknowledgments
 The authors acknowledge insightful conversations with Jorge Rueda, Laura Becerra,  Evan Kirby, Mithi Alexa (Mia) de los Reyes, Lifan Wang, Noam Soker, and Hagai Perets. R.T.F. and S.N. acknowledge support from NASA ATP award 80NSSC18K1013, and K.B. support from NASA XMM-Newton award 80NSSC19K0601. This work used the Extreme Science and Engineering Discovery Environment (XSEDE) Stampede 2 supercomputer at the University of Texas at Austin's Texas Advanced Computing Center through allocation TG-AST100038. XSEDE is supported by National Science Foundation grant number ACI-1548562 \citep{townsetal2014}.


\software { SeBa \citep{toonenetal2012}, FLASH 4.3 \citep{Fryxell_2000, dubeyetal12}, FLASH SN Ia module \citet{townsleyetal16} (\href{http://pages.astronomy.ua.edu/townsley/code}{http://pages.astronomy.ua.edu/townsley/code}), yt  \citep{Turk_2011}, Python programming language \citep{vanrossumetal1991}, Numpy \citep{vanderwaltetal2011}, IPython \citep{perezetal2007}, Matplotlib \citep{hunter2007} }.


\bibliography{sneia}{}
\bibliographystyle{aasjournal}

\end{document}